\documentclass[aip,rsi,reprint,graphicx]{revtex4-1} 

\usepackage{graphicx}
\usepackage{amsmath}
\usepackage{braket}
\usepackage{gensymb}
\usepackage[version=4]{mhchem}

\begin{document}

\title{Fast Reflective Optic-Based Rotational Anisotropy Nonlinear Harmonic Generation Spectrometer}

\author{Baozhu Lu}
\thanks{Baozhu Lu and Jason D. Tran contributed equally to this work}
\affiliation{Department of Physics, Temple University, Philadelphia, PA 19122, USA}
\affiliation{Temple Materials Institute, Temple University, Philadelphia, PA 19122, USA}
\author{Jason D. Tran}
\thanks{Baozhu Lu and Jason D. Tran contributed equally to this work}
\affiliation{Department of Physics, Temple University, Philadelphia, PA 19122, USA}
\affiliation{Temple Materials Institute, Temple University, Philadelphia, PA 19122, USA}
\author{Darius H. Torchinsky}
\email{dtorchin@temple.edu}
\affiliation{Department of Physics, Temple University, Philadelphia, PA 19122, USA}
\affiliation{Temple Materials Institute, Temple University, Philadelphia, PA 19122, USA}

\date{\today}

\begin{abstract}
We present a novel Rotational Anisotropy Nonlinear Harmonic Generation (RA-NHG) apparatus based primarily upon reflective optics. The data acquisition scheme used here allows for fast accumulation of RA-NHG traces, mitigating low frequency noise from laser drift, while permitting real-time adjustment of acquired signals with significantly more data points per unit angle rotation of the optics than other RA-NHG setups. We discuss the design and construction of the optical and electronic components of the device and present example data taken on a GaAs test sample at a variety of wavelengths. The RA-second harmonic generation data for this sample show the expected four-fold rotational symmetry across a broad range of wavelengths, while those for RA-third harmonic generation exhibit evidence of cascaded nonlinear processes possible in acentric crystal structures.
\end{abstract}


\maketitle
\section{Introduction}

Rotational Anisotropy Nonlinear Harmonic Generation (RA-NHG) is an all-optical probe in which the amplitude of the $n^{th}$ harmonic of a laser beam is measured as a function of the rotation of either polarization angle or crystal axes\cite{Fiebig2005,Kirilyuk2005,Mcgilp2010,Denev2011}. In recent years, the technique has found increasing application as a measurement of both lattice symmetry and electronic symmetry breaking states of periodic condensed phase systems\cite{Torchinsky2015,Zhao2016,Harter2017,Padmanabhan2018}, as well as a serving as a sensitive spectroscopic probe of magnetic~\cite{Fiebig2000,Kaminski2010a,Lafrentz2010a,Lafrentz2012,Lafrentz2013a,Lafrentz2013b} semiconducting~\cite{Lim2000,Bergfeld2003} and topological systems\cite{Patankar2018}.

There are two general types of RA-NHG experimental geometry found in the literature: those in which the incident and analyzed polarizations are rotated with a static relationship between the scattering plane and the crystalline axes, and those in which either the scattering plane or the crystal are rotated with a static configuration of incident and outgoing polarizations. While experiments in the former geometry are relatively easy to perform and provide essential information on inversion symmetry breaking or spectroscopic information of nonlinear susceptibility, those in the latter geometry generally access a greater number of nonlinear susceptibility tensor elements per scan and thus provide a more direct picture of the material's inherent symmetries, motivating our focus on this geometry for our device.

Earlier experimental implementations of this geometry consisted of a rotating sample, which presented signficant experimental challenges to alignment and for working in cryogenic environments. Incorporation of a custom-made transmissive binary phase grating, i.e., a phase mask, in conjunction with high numerical aperature imaging optics largely addressed these issues. This configuration allows for oblique incidence upon the sample, defining a scattering plane relative to which the polarization may be described as s-polarized (perpendicular to the scattering plane) or p-polarized (within the scattering plane); as the diffractive optic is rotated, the scattering plane rotates relative to the crystalline axes, mimicking physical rotation of the crystal~\cite{Torchinsky2014}. 

The first diffractive-optic based experimental setup operated in a ``stop-start" manner, wherein a series of optics were repositioned for each acquired angle, taken in $5\degree$ increments. The extremely slow pace of data collection made the setup sensitive to low frequency noise components, most significantly laser drift, that could not be filtered out by lock-in detection. These issues motivated a further significant improvement through a diffractive optic based ``fast-scanning" approach, made possible by spinning a collection of optics at $\sim 5$~Hz using a drive shaft~\cite{Harter2015}. In this configuration, the emitted SHG light is accumulated onto a CCD camera in the form of a spatially dependent pattern that embeds the rotational anisotropy signal $I(\phi)$ which is recorded as a function of rotation angle $\phi$. Numerical data that can be analyzed for symmetry changes are recovered via radial numerical integrals of the acquired image.

One of the main drawbacks of both aforementioned RA-NHG setups is that the phase grating is a highly wavelength-dependent optic. First, the grating's diffraction efficiency is set by the etch depth of the optic and can be inefficient outside of the design wavelength. Additionally, the diffracted beam angle is determined as according to the grating equation $\Lambda = \lambda/2\sin(\theta)$, where $\Lambda$ is the grating feature size, $\lambda$ is the wavelength of light, and $\theta$ is the angle subtended relative to the surface of the grating. This strong wavelength dependence presents a problem for spectroscopic studies that could reveal magnetic\cite{Fiebig2000, Kaminski2010a, Lafrentz2010a, Lafrentz2012, Lafrentz2013a, Lafrentz2013b}  or electronic\cite{Bergfeld2003,Patankar2018,Mund2018} resonances in condensed phase systems. As multiple experimental parameters involving alignment change with $\lambda$, the combination of different tensor elements of interest will change as a function of angle of incidence for p-polarized incoming or outgoing beams, obscuring their intrinsic energy-dependence. This drawback has meant that truly spectroscopic studies have been done exclusively in a static scattering plane geometry.

Here we describe a novel design for a fast RA-NHG spectrometer that requires neither custom made diffractive optics nor CCD camera detection and allows for continuous wavelength tuning so that spectral features can be accurately resolved. The system is built upon lockstepped motion of a stepper motor driven polarizer and a voice-coil fast turning mirror as synchronized by a data acquisition card and stepper motor controller. Owing to the predominance of reflective optics in its construction, the setup can be achromatized. Being driven by precisely timed controllers, the spectrometer desribed here can be interfaced directly with more sensitive RA-SHG measurement techniques~\cite{Lu2018a}.

This paper is organized as follows. In Section~\ref{sec:des}, we describe the principle behind the experimental construction, providing detailed layouts of the optics and the system electronics. In Section~\ref{sec:align}, we provide a detailed alignment procedure we have developed to ensure proper functioning of the apparatus. In Section~\ref{sec:randd}, we present an example of the use of the device to examine the RA-SHG signal of GaAs at multiple wavelengths and in Section~\ref{sec:conc} we conclude by summarizing the essential points of the paper and providing directions for further improvements.

\section{Experimental Design}\label{sec:des}

\subsection{Optics}

\begin{figure}
    \includegraphics[width=0.48\textwidth]{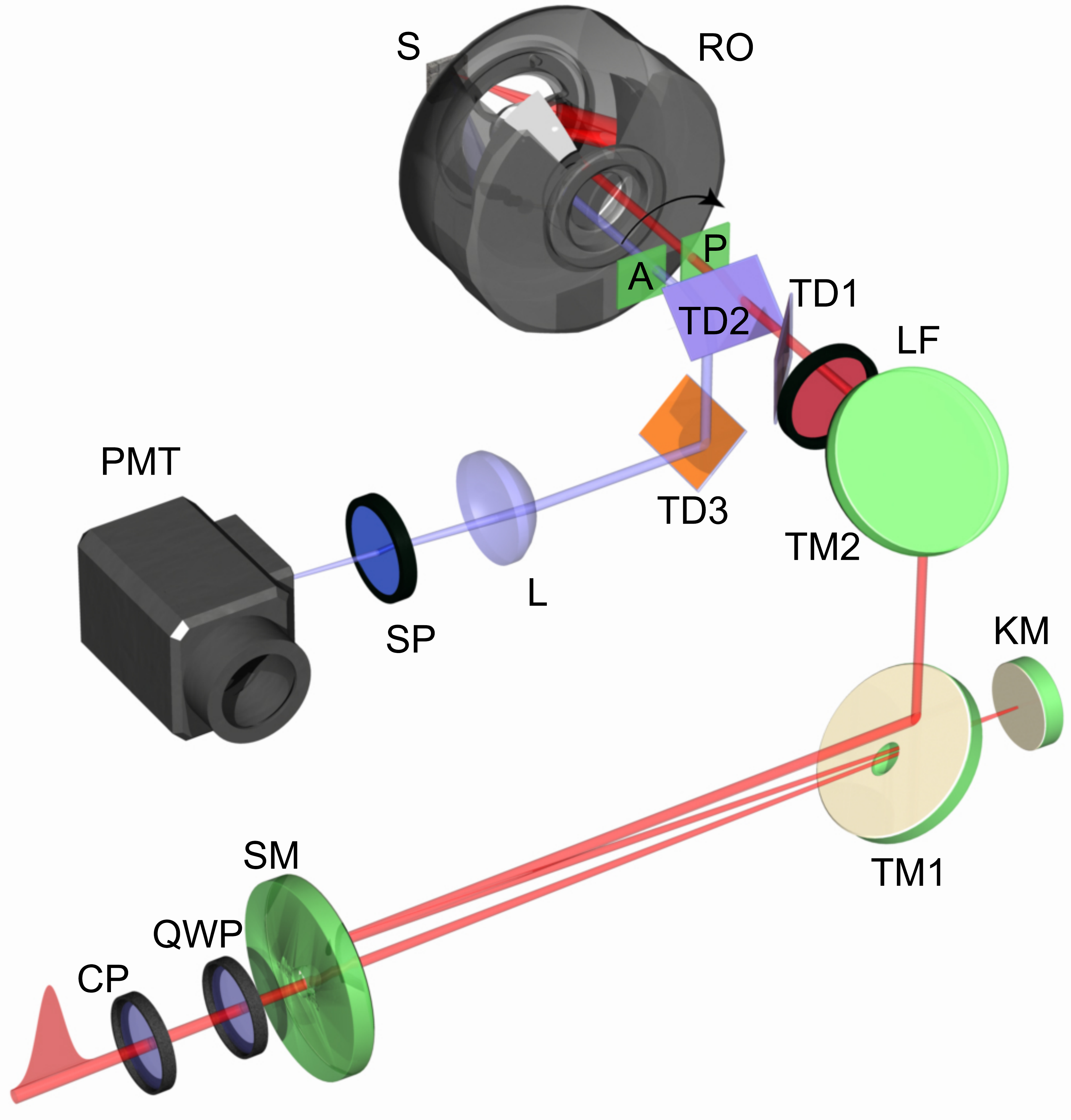}
    \caption{Perspective diagram of the optical configuration of the RA-SHG spectrometer showing all elements. These are, in order of first incidence: CP - cleanup polarizer, QWP - quarter wave plate, KM - kinetic mirror, SM - spherical mirror, TM1 - first periscope mirror, TM2 - second periscope mirror, LF - group of longpass filters, TD1 - dichroic one, P - polarizer, RO - reflective obective, S - sample, A - analyzer, TD2 - dichroic two, TD3 - dichroic three, L - lens, SF - grouping of shortpass and/or bandpass filters. Not shown is the optic used to focus the beam into the setup.}
    \label{fig:perspective}
\end{figure}

\begin{figure}
    \includegraphics[width=0.48\textwidth]{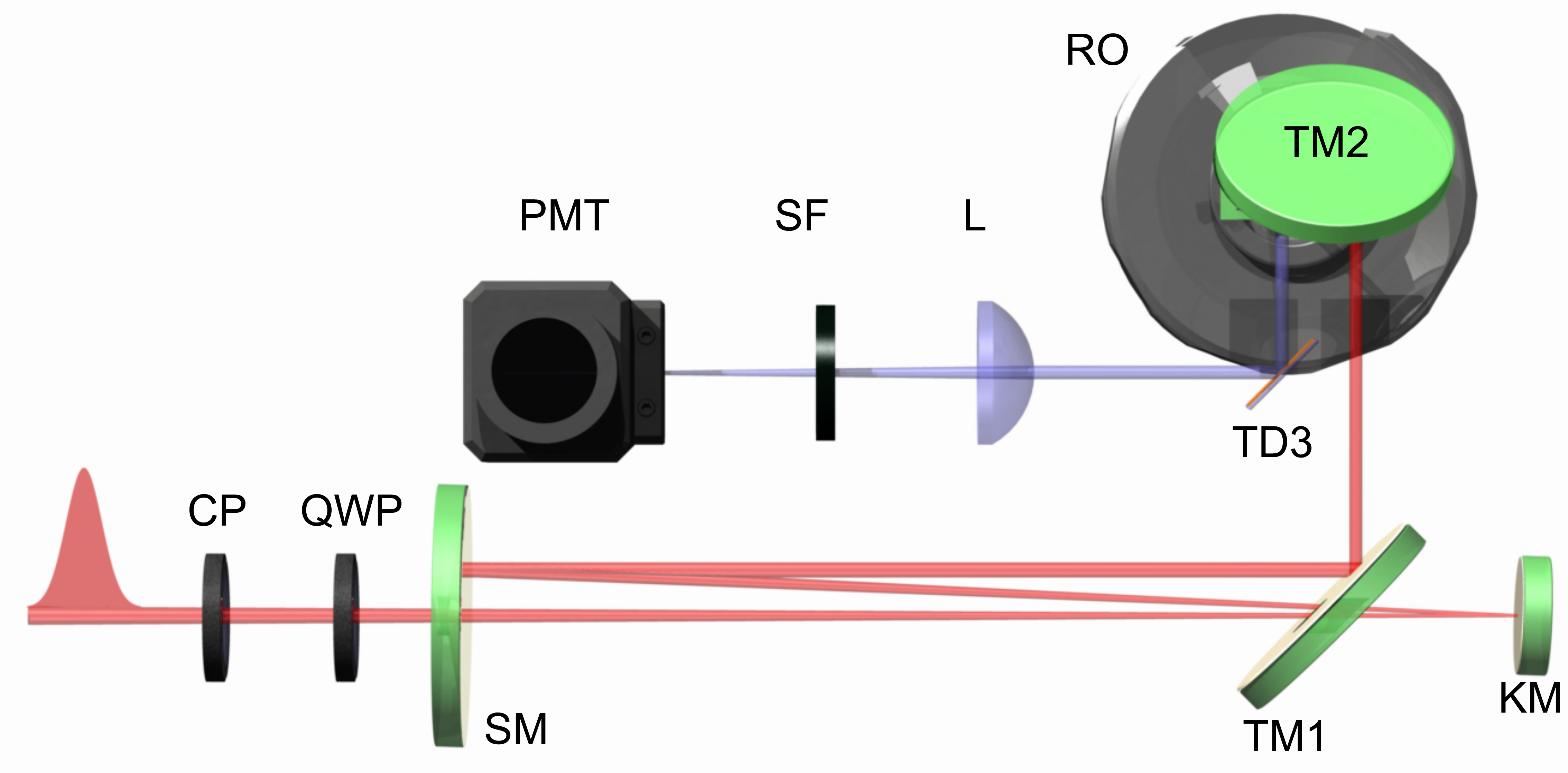}
    \caption{Side view of the RA-SHG apparatus. Note that the incoming beam passes through holes in SM and TM1 before being incident on KM. The bore of the hole in TM1 is canted such that it is parallel to the incoming beam path when TM1 is in place at a $45\degree$ angle. The mirror KM here is depicted laterally further away from TM1 than it is in practice for clarity. The tilt angle of KM was chosen to be such that it would not clip on either the bores of TM1 and SM or on the entrance aperture of RO.}
    \label{fig:side}
\end{figure}

\begin{figure}
    \includegraphics[width=0.48\textwidth]{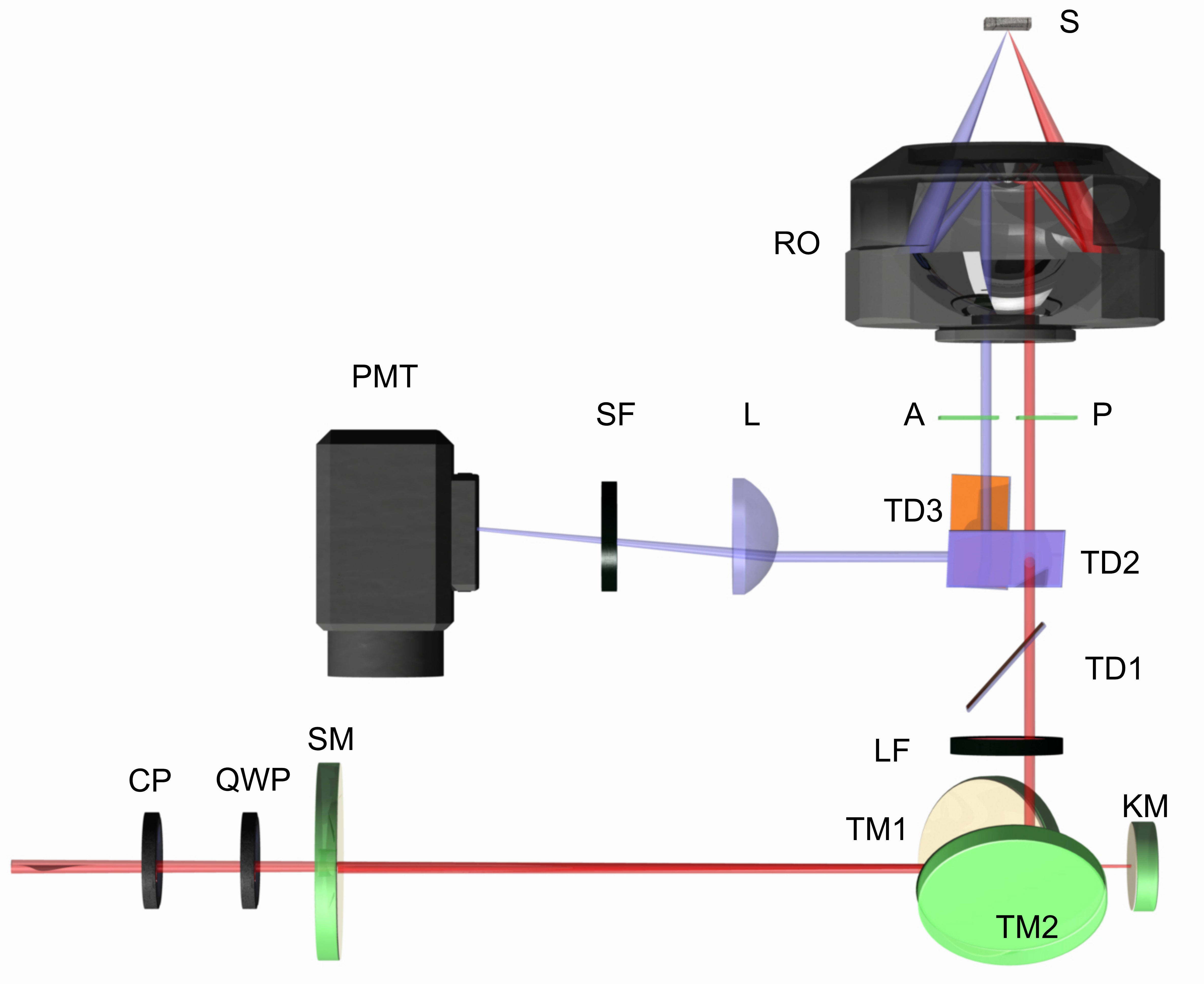}
    \caption{Top-down view of the RA-SHG spectrometer. Note that the beam is reflected upwards by KM, as seen in Fig.~\ref{fig:side}, but due to the right-angle periscope, emerges to the righthand side of the RO entrance aperture. The sense of rotation of the kinetic mirror and the scattering plane are thus offset by $90\degree$.}
    \label{fig:top}
\end{figure}

The RA-SHG setup we have developed is depicted in Figs.~\ref{fig:perspective}, \ref{fig:side} and \ref{fig:top}. Comprising predominantly of reflecting optics, the apparatus was designed to be used with a tunable laser source, such as an optical parametric amplifier (OPA), as was used in the data collection below. The OPA output beam passes through multiple short and long pass filters chosen to remove various parasitic signal contributions from the OPA's colinear second stage. Although shown in the Figures as focusing on the kinetic mirror (KM) for clarity, the beam is focused by a 50~cm lens to a point immediately before the reflective objective (RO) so that it may be approximately collimated and avoid excessive SHG contributions from longitudinal field components~\cite{Quesnel1998,Carrasco2006}. As the wavelength is changed, the shift in this focal position as a function of wavelength is minimal due to its inherently large Rayleigh length. We note that the lens can easily be replaced by a reflecting converging mirror if an extremely wide wavelength range is used.

After the lens (not shown), the OPA output passes through a polarizer (CP) to purify the polarization state and then a quarter wave plate (QWP) to convert the polarization from linear to circular. It then passes through a custom 4.92~mm hole bored in the middle of a 20" focal length protected aluminum spherical mirror (SM - Edmund Optics 32-821). The beam then passes through the middle of a 4.92~mm hole bored at a $45\degree$ angle in a tilted turning mirror (TM1 - Edmund Optics 68-333), after which it is incident upon the center of the kinetic mirror (KM). KM comprises a standard silver mirror (Newport 10D20ER.2) seated upon a voice coil motor set (Optics in Motion OIM5001) which is able to tilt at high speeds to deflect the beam so that it may ultimately be introduced at oblique incidence on a sample. We note that KM plays the same role as the phase mask in Refs.~\onlinecite{Torchinsky2014,Harter2015} in that it introduces an oblique angle by reflection instead of by diffraction, with the tilt angle proportional to input control voltage in the X and Y direction by $1\degree/$V. The KM is capable of extremely fast motion with a 3dB point in the conversion of voltage to angle at $\approx 430$~Hz with accompanying shift in phase of $-100\degree$. Both the attenuation of the tilt angle and the phase delay are negligible at the $5$~Hz frequency used here.

The KM reflects the beam at an angle chosen such that it may pass back through the bore in the center of TM1 but not through the hole in SM. It is essential that the surface of KM lies in the focal plane of SM so that reflection off of SM brings the beam parallel to, but offset from, the incoming optical axis. The beam then reflects off of the two turning mirrors TM1 and TM2 (Thorlabs PF20-03-P01), offset from the hole in TM1. The turning mirrors are in a 90-degree rotated periscope configuration chosen to compensate for reflectivity anisotropy between the s- and p-polarized light. Since the beam is at near-normal incidence for both KM and SM, there is little p-polarization component in the beam and s/p asymmetry in these optics can be ignored.

After the beam is reflected by TM2, parasitic nonlinear harmonic light is removed by long pass filtering (LF - Various Thorlabs FELH and FESH series optics) before passing through a triple-dichroic group (TD1-3 - Various seated in a Thorlabs CM1-DCH mount, Thorlabs DMLP550R used below) that has equal amount of s-polarization and p-polarization reflectivity and transmissivity, maintaining a circular polarization for the OPA field upon its emergence. In principle, the use of a triple dichroic is unnecessary for the wavelengths where the s/p asymmetry in both reflection and transmission is minimal. In particular for the DMLP550R this falls at the native Ti:sapphire wavelength of 800~nm.

Next, the laser pulse passes through a spinning broadband wire-grid polarizer (Thorlabs - WP12L-UB) to determine the incoming polarization as either in (p-) or perpendicular to (s-) the scattering plane. The polarizer is seated in a custom made mount attached to the spinning stepper motor stage (Standa 8MRU-1) that has had its front plate removed. The beam is then delivered off-center of a NA=0.5 reflective objective (Edmund Optics - 68-188), which brings it in at $\sim 26\degree$ angle of incidence, allowing for ready discrimination between s- and p-polarized light. As described above, the beam is roughly collimated by the RO primarily to avoid parasitic contributions to the NHG as tightly focused beams contain much larger longitudinal electric field components and thus constitute a large departure from the ideal plane wave for which RA-NHG signals are generally calculated. Furthermore, producing a larger spot on the sample also allows more integrated power to be used to achieve the same fluence, which in turn increases the integrated emitted signal strength.

After reflecting off the sample (S) and reemerging from the RO, the second harmonic component of the beam passes through the analyzer A (Thorlabs - WP12L-UB) which is in the same rotation stage as the analyzer P, selecting the outgoing polarization as either S- or P-polarized depending upon how it is set by the experimentalist. The SHG pulse then reflects from the triple dichroic mirror stack toward the detection arm. A 7.5~cm lens (L - Thorlabs LA4380-A) then focuses the beam down past two shortpass filters and one bandpass filter (grouped in SF - various, Thorlabs FESH550 and FBH400-40 used below) to remove the fundamental before the amplitude is detected by a photomultiplier tube (PMT - Hamamatsu R12829). 

\subsection{Electronics}

The PMT used here is biased by a high voltage power supply socket assembly (Hamamatsu - C12597-01). Upon excitation by incident SHG photons, it emits a short current pulse proportional to the light intensity through conversion by a factor of roughly $4\times 10^4$~A/W or, more appropriately for pulsed light, $4\times 10^4$~C/J. The PMT output was connected to the input of a charge integrator (Cremat - CR-Z-PMT) with a 1.3~mV/pC conversion and then converted into a voltage pulse by a shaper instrument (Cremat - CR-S-8us-US) with the amplification factor set to $10\times$. Ignoring the quantum efficiency of the PMT, the PMT bias voltages used here provided a conversion efficiency of $260$~$\mu$V/emitted photon at a photon energy of 3.0~eV. The overall sensitivity is determined by the amplification and dark current of the photomultiplier while the electronic noise floor is determined by the amplification from the integrator and shaper.

\begin{figure}
    \includegraphics[width=0.48\textwidth]{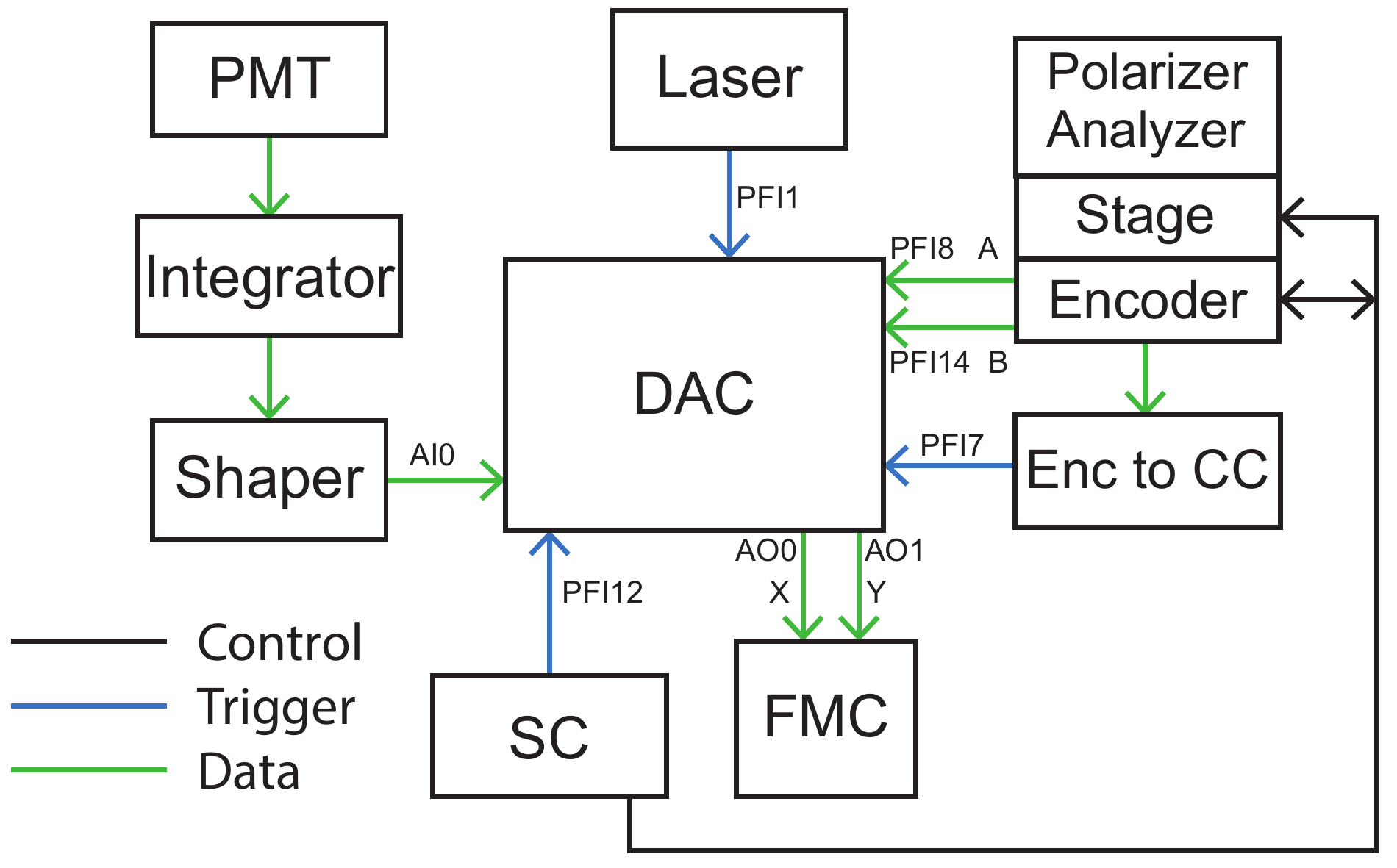}
    \caption{Schematic diagram of electrical connections in the RA-SHG spectrometer. Abbreviations are: DAC - data acquisition card, PMT - photomultiplier tube, An. Stage - Analyzer rotation stage, Pol - Polarizer rotation stage, SC - stage controller, Enc to CC - Encoder to counter converter, FMC - Fast mirror controller. The various other labels (AI0, PFI12, etc.) indicate input and output connections on the DAC.}
    \label{fig:elec}
\end{figure}

\begin{figure*}
    \includegraphics[width=\textwidth]{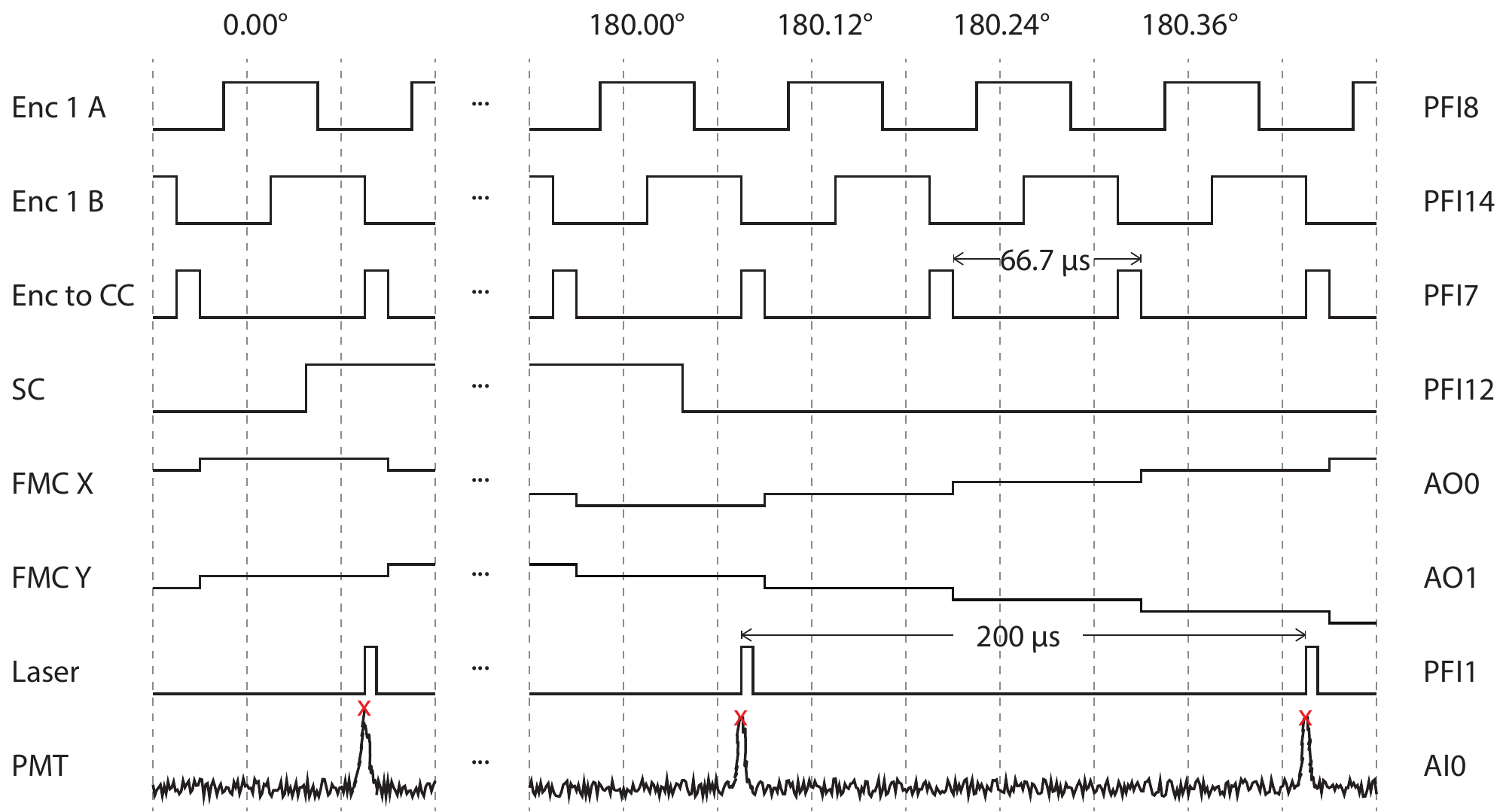}
    \caption{Timing diagram for the electronics depicted in Figure~\ref{fig:elec}. The DAC is triggered to start acquisition every $360\degree$ by SC which is taken from Zaber and input into PFI12 of the DAC. The ENC A and ENC B inputs represent the quadrature encoder monitoring the stage rotation through 3000 pulses per revolution connected to the PFI8 and PFI14 inputs of the DAC, respectively. ENC to CC is the output counter pulse that triggers once every time the encoder has detected rotation by $0.12\degree$ as read though DAC channel PFI7. The instantaneous voltage driving the X (Y) coordinate of the FM is shown as FMC X (FMC Y) output from DAC channel AO0  (AO1) and representing the function $X=A\cos(\phi)+X_0$ ($Y=A\sin(\phi)+Y_0$). A laser-synchronized trigger on DAC channel PFI1 records the output of the detection electronics on DAC input AI0. Although shown here as being synchronous with the stage rotation, it was asynchronous in our experiments.}
    \label{fig:timing}
\end{figure*}

Obtaining a full $360\degree$ RA-SHG signal requires accurate timing between the kinetic components (KM, P and A) as well as synchronization with the laser output. A diagram of the electrical connections is shown in Fig.~\ref{fig:elec} while a timing diagram is shown in Fig.~\ref{fig:timing}. All components communicated with a data-acquisition card (DAC - NI USB-6281) that simultaneously read the instantaneous angle of the polarizer, directed the tilt angle of the kinetic mirror and recorded the voltage value of the SHG from the PMT. The DAC was programmed in the LabView environment.

The polarizer/analyzer stage is rotated by software to starting position and then spun at 300~RPM by a stepper motor controller (Zaber X-MCB2-KX12B). We were able to drive the stage rotation speed to 600~RPM (i.e,, 10~Hz) with minimal problems, and could in principle push it significantly higher with the use of servo motors at the expense of the number of data points per trace. The stage is equipped with a digital encoder (Broadcom HEDM-5500{\#}B14) that provides 1000 counts per revolution of the stepper motor. Due to the 3:1 gearing ratio between the drive and the stage mount, this equates to 3000 counts per revolution of the spinning optic and in principle gives the experiment a $0.12\degree$ angular resolution provided that data can be acquired at 15~kHz. While the encoders is output to the stepper controller to assist in closed-loop motion, only the A and B encoder outputs are needed to synchronize the spinning of the stage with the KM as described below. These encoder outputs are connected to counter inputs on the DAC card (DAC channels PFI8 and PFI14, respectively) to provide an instantaneous count read of the number of $0.12\degree$ steps advanced by the stage. The stage spins at a constant rate for the duration of the measurement, set in software as a predetermined number of revolutions depending upon how many averages are needed to provide usable data. The stage controller is simultaneously programmed to provide a positive edge TTL trigger output to the DAC card exactly when the stages reach $0\degree$ (PFI12) to indicate to the DAC to send the data to the computer and begin the next $360\degree$ acquisition.

The voice-coil KM is controlled by two analog voltage inputs on X and Y channels (DAC channels AO0 and AO1, respectively) to tilt the mirror by 0.5$\degree$/V in the X or Y directions, respectively. In order to generate the requisite obliquely incident beam, the signals on the two channels must be given by $\textrm{X}=A\cos\left(\phi\right)+X_0$ and $\textrm{Y}=A\sin\left(\phi\right)+Y_0$ where the angle $\phi$ is identical to the rotation angle of the stages from their starting positions, $A$ is the amplitude of the deflection in volts, and $X_0, Y_0$ are offsets determined in order to make the beam retroreflect upon its input path when $A=0$, which is essential to ensure that the beam is centered about the optical axis.

As the encoder resolution is 3000 steps per revolution of the stages, we sampled the functions $\textrm{X}=A\cos\left(\phi\right)+X_0$ and $\textrm{Y}=A\sin\left(\phi\right)+Y_0$ by 3000 equally spaced angular points that were uploaded into the regeneratively output onboard memory of the DAC. As the stages rotate, each encoder pulse pair from its A and B outputs is converted into a counter pulse by an encoder to counter interface board (US Digital PC6-C-1-H5) providing a single TTL output (DAC channel PFI7). This TTL pulse triggers the DAC to output the next sample of the X and Y signals on the AO0 and AO1 analog channels represented as {FMC X} and {FMC Y} in Fig.~\ref{fig:timing}. Each subsequent point in the sampled waveform is thus triggered by the encoder detecting movement by one count. In this way, the voice coil-driven mirror is perfectly synchronized with the fast stages. We emphasize that the encoder signals must be converted into a counter signal. If an encoder output is directly used as a trigger, the resulting improper counting of pulses/revolution introduces a continuously worsening phase slip between kinetic mirror and stages. Given the sufficiently high bandwidth of the kinetic mirror and the small sampling step of the mirror drive voltage, we phase slip between the synchronization between the stages and mirror was immeasurable.

The output of the digital delay generator (DDG) electronics timing the Pockels Cell of the regen cavity provides an appropriate TTL signal that acts as a data acquisition trigger for the analog input of the DAC (Laser in Figs~\ref{fig:elec} and \ref{fig:timing}). Every such pulse triggers the DAC to simultaneously sample the shaper output at its maximum and record the angular position of the encoder. We note that timed sampling significantly reduces spurious signals as may arise from incoherent stray light, cosmic rays or dark current from the PMT since the data were only sampled over the sampling period of the DAC as compared with the 200~$\mu$s repetition period of the measurement, resulting in a low duty cycle measurement and commensurate reduction in noise. Due to the $\sim 5$~Hz speed of the stages, 1007 data points are taken per revolution (503 points per revolution are acquired at $\sim 10$~Hz), and every voltage acquisition is assigned a unique angular step. Every full revolution, the DAC uploads the data points to a computer that records a running average of the data. Both Cartesian and polar plots of the last $360\degree$ sweep and the running average are simultaneously plotted to allow real-time fine tuning of the apparatus during alignment and to monitor the signal during data acquisition.

We note that the laser and the stages do not need to be synchronized. Within one revolution of the stages, there are 3000 identifiable angular steps and, at these repetition rates, the possibility of $\sim 1000$ data points. Every time the laser fires, both the SHG amplitude and the angle are simultaneously measured and averaged with the previous sweep. Thus, each data point represents an average voltage for an average angle within an angular range of $\sim \pm 0.18\degree$.

\section{Alignment Procedure}\label{sec:align}

Here we describe a detailed alignment procedure. After roughly positioning all the optics, the focusing lens is placed in the path of the incoming beam and the focus fixed roughly an inch in front of the RO. Then, the focusing beam is introduced along the central optical axis of the bottom tier of optics of Fig.~\ref{fig:side}, taking care to ensure that the beam is centered on the bores of the mirrors SM and TM1. This step is assisted by mounting optics CP, QWP, SM, TM1, TM2, KM and LF in a 60~mm cage system, and using tools that allow for accurate alignment of the beam down the central axis. Once this has been done as precisely as possible, the KM is adjusted so that the beam is centered on the reflective surface.

Next, constant voltage offsets are applied to the X and Y inputs of the KM controller using the DAC in order to retroreflect the beam. These values are tuned to within 10~mV, following the beam as far back as possible. The offset values are used to drive the KM according to the function $\textrm{X}=A\cos\left(\phi\right)+X_0$ and $\textrm{Y}=A\sin\left(\phi\right)+Y_0$ where the value of $A$ is chosen such that the ring described by the beam is approximately 9~mm. This diameter is small enough such that the beam is not clipped on the bore of TM1 upon reflection off of the KM, but not so small that it would make it through the bore of the SM or be obscured by the central reflector of the RO.

The KM is then driven regeneratively by the DAC at a fast enough rate so that the exiting beam appears as a ring upon reflection from SM and is made to traverse the lower tier parallel to the optical axis. Next, the angles of TM1 and TM2 are adjusted so that the ring is directed directly upwards and then at a right angle to the lower tier of optics and parallel with the optical table. 

We determined that the incoming polarization is as close to circular as possible by placing a photodiode after the spinning polarizer (optic P in Figs.~\ref{fig:perspective}-\ref{fig:top}) and sending the output into the sampling system. A perfectly circularly polarized beam results in a round, featureless output. This was achieved by gradually adjusting the QWP both by rotation and by tilting about its axis until as flat of a line was observed as possible. In practice, we were able to recover a flatness with peak-to-peak deviation 0.9\% routinely.

The sample is aligned by stopping the KM such that the laser is at a single point and making the beam retroreflect off of it. The KM is restarted and the sample is translated to the center of the ring. In our apparatus, the RO is mounted in a rotatable tip/tilt mount (Thorlabs KS1RS). Before the RO is placed into the setup, a mirror inside of a lens tube is inserted into the mount and the mount adjusted to retroreflect the beam. Then, a target is placed in the mount in place of the lens tube and the mount translated on an XYZ translation stage to center the ring on the target. These steps ensure centration and normal incidence of the RO.

Once the RO is placed in the setup, the sample is adjusted on an XYZ stage to bring it to the center of the beam, and the beam made to reflect straight back upon itself. We then checked that the beam would be able to only hit one point on the sample for the full $360\degree$ revolution of the experiment. In principle, this is accomplished by making sure that the beam emerging from the turning mirror TM2 describes a circle centered on the optical axis of the upper tier of optics in Fig.~\ref{fig:side}. More careful adjustment of this step is accomplished by using a pellicle beamsplitter (Thorlabs BP250) to image the surface of the sample focusing the beam emerging from the RO onto a bare CCD (Logitech C170 Webcam with the optics removed) using a 40~cm lens (Thorlabs AC508-400-B), i.e., through building a basic microscope. As the surface is imaged, fine tuning adjustment of the sample plane is made to ensure that the beam walks on the sample as little as possible. 

The triple-dichroic stack is then placed such that is center matches that of the ring and the mirror sits at $45\degree$ relative to the incoming beam. Although rated to transmit the fundamental, enough parasitic light is often present for the ring to be imaged by an IR viewer and centered on the position where the active area of the PMT normally sits. The lens L is adjusted to make sure that the beam is as stationary as possible at this point, after which point the remaining optics are placed in the beam (A and SF).

\section{Results and Discussion}\label{sec:randd}

\ce{GaAs} was chosen a test sample due to widespread availability of high quality single crystals, as well as due to a relatively large second harmonic generation (SHG) susceptibility $\chi^{(2)}_{ijk}$. The expected RA-SHG signals as a function of scattering plane rotation $\phi$ for a (001) oriented sample are given by
\begin{align}
I^{SS}_{2\omega}(\phi)&= 0\\
I^{PS}_{2\omega}(\phi) &= 2\chi^2_{xyz}\left(1+\cos\left(4\phi\right)\right)\label{eq:SHG}\\
I^{SP}_{2\omega}(\phi) &= \chi^2_{xyz}/2\left(1-\cos\left(4\phi\right)\right)\\
I^{PP}_{2\omega}(\phi) &= \chi^2_{xyz}/2\left(1-\cos\left(4\phi\right)\right).
\end{align}
where the superscript indicates the polarization of the incoming and outgoing beams in succession, i.e., $I^{PS}_{2\omega}(\phi)$ specifies incoming light with polarization vector in the plane of incidence and outgoing light with polarization vector perpendicular to the plane of incidence. The PS geometry was chosen for experiments here since other responses are shown to be strongly influenced by electric field induced second harmonic (EFISH) from surface charging~\cite{Germer1997}. Consequently, SHG also proceeds at third order with one of the field components being the zero frequency electric field perpendicular to the surface; the resulting EFISH heterodynes with the direct SHG and yields asymmetric RA-SHG traces, least notably for the PS geometry.

\begin{figure}
    \includegraphics[width=0.48\textwidth]{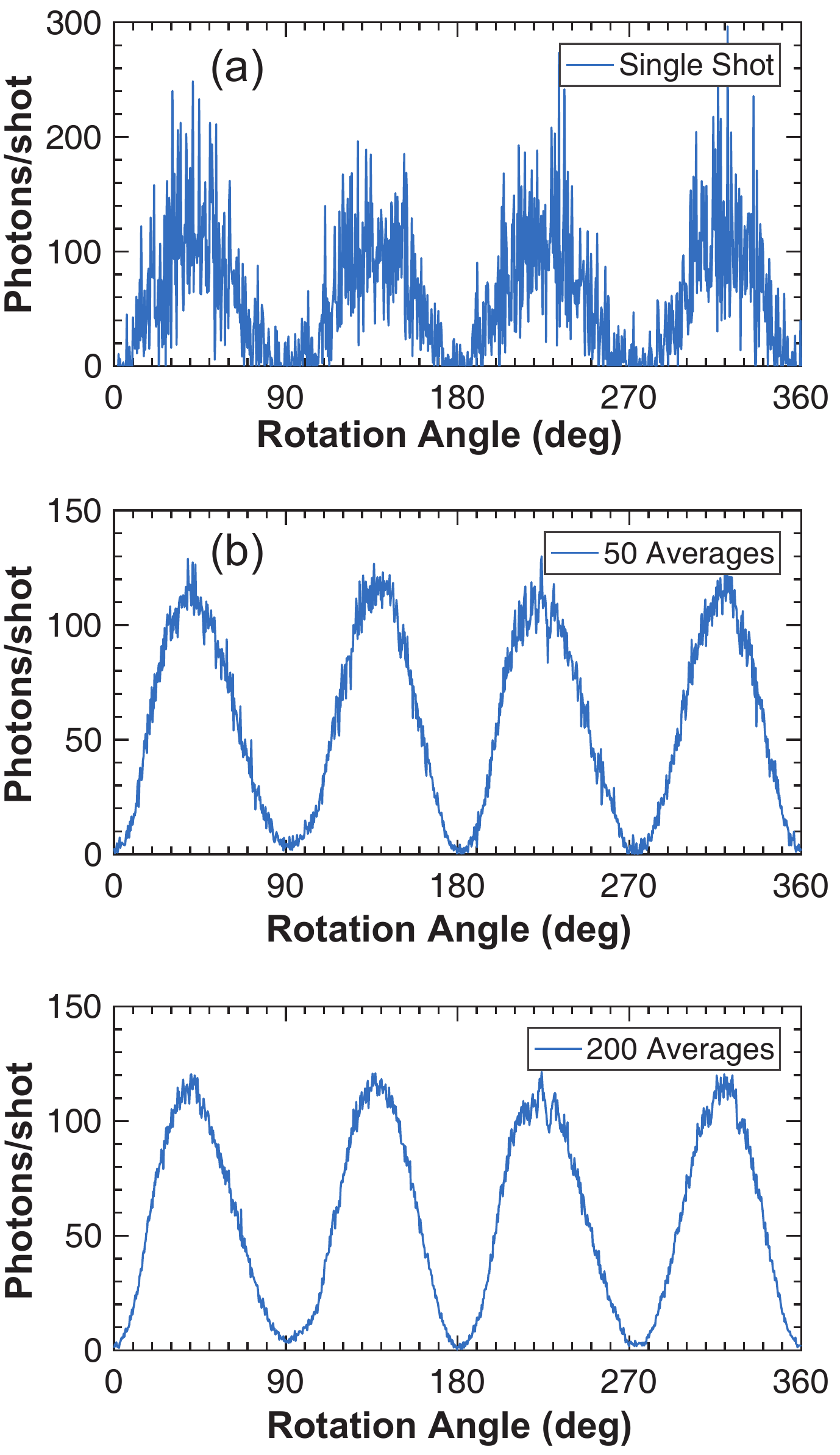}
    \caption{Data on \ce{GaAs} taken in the PS experimental geometry with different degrees of averaging, including (a) single shot (i.e., one circulation of the optics) (b) 50 averages and (c) 200 averages. Each average takes 200~ms.}
    \label{fig:data1}
\end{figure}

We also recorded third harmonic generation (THG) signals. For the \ce{GaAs} test sample, the RA-THG signals were computed to be
\begin{multline}
I^{SS}_{3\omega}(\phi)= \left(\frac{3}{4}(\chi_{xxxx}+\chi_{xxyy})\right. \\+ \left.\frac{1}{4}(\chi_{xxxx}-3\chi_{xxyy})\cos(4\phi)\right)^2
\end{multline}
\begin{align}
I^{PS}_{3\omega}(\phi) &= \left((\chi_{xxxx} - 3\chi_{xxyy})\sin(4\phi)/4\right)^2\label{eq:thgps}\\
I^{SP}_{3\omega}(\phi) &= \left((\chi_{xxxx} - 3\chi_{xxyy})\sin(4\phi)/4\right)^2
\end{align}
\begin{multline}
I^{PP}_{3\omega}(\phi) = (\chi_{xxxx} + 3\chi_{xxyy})^2 +  \left(\frac{3}{4}\chi_{xxxx} + \frac{15}{4}\chi_{xxyy}\right.\\ + \left.\frac{1}{4}(\chi_{xxxx} - 3\chi_{xxyy})\cos(4\phi))/4\right)^2.
\end{multline}

We used the output of a Coherent Astrella regeneratively amplified Ti:sapphire laser lasing at 800~nm at 5~kHz repetition rate and producing 35~fs pulses to seed the input of an Optical Parametric Amplier (OPA - Light Conversion TOPAS TWIN). The OPA produced tunable laser pulses from 480~nm to 2.6~$\mu$m. In this study, we focused on the incoming wavelength range from 800~nm to 1.3~$\mu$m. Significantly longer wavelengths would produce second harmonic photons below bandgap in \ce{GaAs}, which and thus need to be measured in transmission, while significantly shorter ones would generate harmonics at wavelengths not easily detected by the PMT.

In Fig.~\ref{fig:data1}, we show RA-SHG data for 800~nm light in/400~nm light out using the PS-polarization combination for various degrees of signal averaging. The dichroics used were Thorlabs DMLP550 longpass dichroic mirrors as TD1-3, and the filters in SF were 3 Thorlabs FBH400-40 bandpass filters with two Thorlabs FESH0500 shortpass filters and one Thorlabs FESH0600 filter. As the optics rotate at 5~Hz, the data are acquired every 200~ms, and recorded as the ``signal shot'' data shown in panel (a). Even though there are, on average, only $\sim150$ photons/shot at the laser fluences used here, we are still able to perceive the four-fold symmetry described by Eq.~\ref{eq:SHG}. We have found that having real-time, quickly updated scans available is extremely useful for the fine tuning experimental alignment, and a unique characteristic of this apparatus. Also shown in Fig.~\ref{fig:data1}(b) and (c) are progressively larger degrees of averaging. After 50 averages, which takes $\sim 10$~seconds, the data have become significantly more clean with only 120 photons/shot on average at the maximum. The additional 150 averages, as shown in panel (c) of the Figure, yield only marginal benefit in signal-to-noise ratio.

\begin{figure*}
    \includegraphics[width=\textwidth]{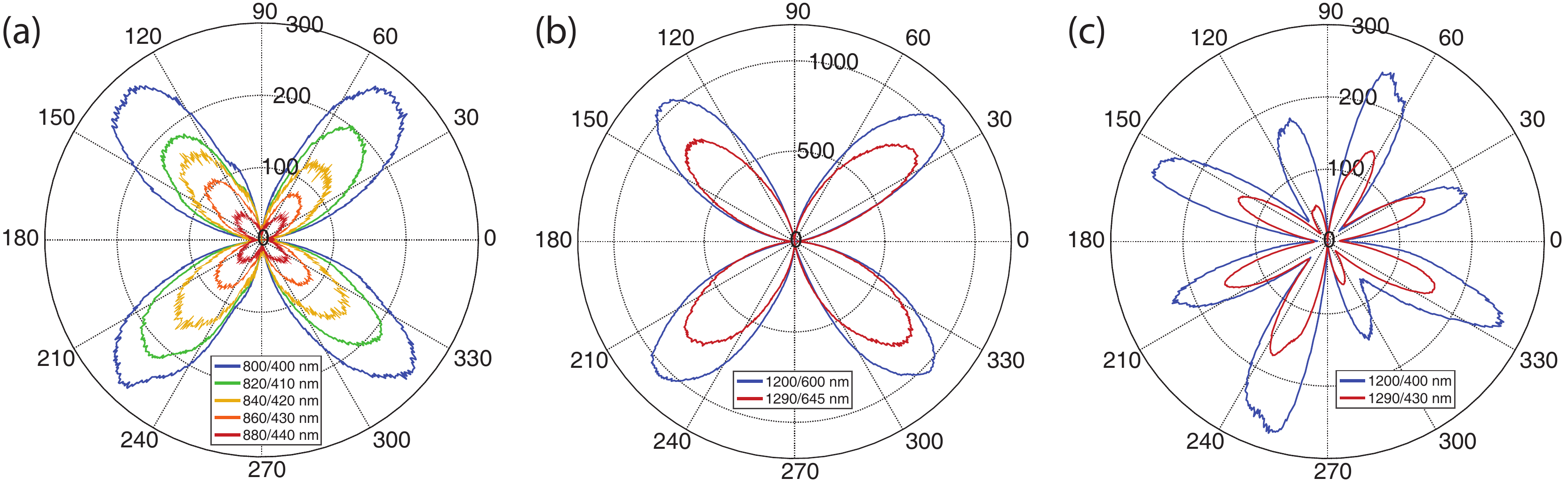}
    \caption{Wavelength dependent polar plots of RA-NHG data on \ce{GaAs}. In all panels, the r-dimension is in units of photons/shot, and in all panels the data from the shortest incoming wavelength hasn't been scaled while all other traces have. (a) RA-SHG data taken every 20~nm from 800~nm to 880~nm incident wavelength. (b) RA-SHG data taken at 1200~nm and 1290~nm using the same setup. (c) RA-THG data in \ce{GaAs} taken at 1200~nm and 1290~nm obtained simply by changing the filters on the PMT from the configuration used for panel (b). The clear departure from the expected 8-fold symmetric pattern for RA-THG is likely due to interference between direct THG and that generated through a cascade of SHG to SFG.}
    \label{fig:polar}
\end{figure*}

Next, we demonstrate the functionality of the apparatus at different incoming wavelengths. Fig.~\ref{fig:polar}(a) shows data from 800~nm-880~nm acquired in steps of 20~nm. The filters in SF from 820~nm - 880~nm are two each of Thorlabs FESH0500 and FESH0650 shortpass filters and one Thorlabs FESH0650 shortpass filter. The data have been scaled by arbitrary factors so that they can all be represented in the same plot. However, the data at 800~nm have not been scaled, and thus yield $\sim 260$~photons/shot on average at the peaks. We observe that there is no change in symmetry in the plot as a function of wavelength. We note that data acquisition of one trace to the next was performed without any change in alignment whatsoever, and thus proceeded very rapidly, i.e., in a few minutes.

Extending the range of data acquisition to longer incident wavelength, RA-SHG data recorded with incoming $\lambda=1200$~nm and 1290~nm are shown in Fig.~\ref{fig:polar}(b). The dichroic mirrors TD1-3 were Thorlabs DMLP900 longpass dichroic mirrors, while the filters used as SF for 1200 nm incident light were one Thorlabs FB600-40 bandpass filter with two Thorlabs FESH650 filters and one Thorlabs FESH0700 shortpass filter. The filters used as SF for 1290~nm incident light were one Thorlabs FB650-40 bandpass filter with two Thorlabs FESH700 shortpass filters. This wavelength range is below the 1.424~eV bandgap and thus higher laser fluences could be applied to the sample before damaging it, yielding $\sim 10\times$ larger signals. Again, the longer wavelength data in the panel have been scaled by an arbitrary relative factor so that they can be easily discerned from one another on the same plot. The data are of exceptional quality, and again show the characteristic four-fold rotational anisotropy expected for SHG from \ce{GaAs} in the PS geometry.

Finally, we show RA-THG data obtained at 1200~nm and 1290~nm. The dichroic mirrors used were the same as for RA-SHG. The filters used as SF for 1200~nm were three Thorlabs FBH400-40 bandpass filters, two Thorlabs FESH0500 shortpass filters and one Thorlabs FESH0600 shortpass filter while those for 1290~nm were one Thorlabs FB430-10 bandpass filter with two Thorlabs FESH0650 shortpass filters and one FESH0600 shortpass filter. We note that the traces do not match the expected pure eight-fold symmetry as derived from the simple calculation in Eq.~\ref{eq:thgps}. We posit that, in common with the RA-SHG traces for geometries not shown here (i.e., SP and PP), this is due to the presence of heterodyning signal channels. Whereas in SHG, contributions can arise from EFISH as described above, THG in non-centrosymmetric materials allows for a parallel but significant two-step pathway for photons to be converted from $\omega$ to $3\omega$ known as a cascade process, whereby by a second harmonic photon mixes with a fundamental photon through sum-frequency generation~\cite{Meredith1981, Bosshard2000}. The rotational anisotropy of this cascaded process is considerably more complicated than is accounted for by the model that yielded Eq.~\ref{eq:thgps}. As can be seen in Fig.~\ref{fig:polar}(c), there is a noticeable wavelength dependence to this effect which will not be a focus of the discussion here.

\section{Conclusions}\label{sec:conc}

We have described the design, alignment and implementation of a fast, reflective rotational anisotropy nonlinear harmonic generation spectrometer and demonstrated its ability to measure RA-NHG signals at various wavelengths, for both SHG and THG responses. The spectrometer provides data of exceptional quality, even down to the regime $\sim 100$~photons/laser shot, allowing it to be fine-tuned in real time by using visual feedback of the signal. The primarily reflective optic composition of the device allows for spectroscopic applications to be performed relatively quickly, with the primary speed bottlenecks arising from the collection of filters and dichroics that must be continuously changed to block the residual fundamental and parasitic wavelengths from being measured by the detector. Notably, this does not change any of the more crucially aligned elements, implying that data acquisition of RA-SHG traces can proceed in a relatively expedient manner and with arbitrarily small or large changes in incident wavelength. However, data acquisition could be further streamlined by replacing the triple-dichroic stack by a spinning thin silver mirror angled to reflect the NHG beam provided a rotary stage that is either thin enough (or of large enough aperture) could be found to allow for the incoming beam to not be occluded over the full rotation of the optics. 

Although not demonstrated here, it would be possible to perform non-degenerate nonlinear optical rotational anisotropy experiments, such as sum-frequency or difference-frequency generation by combining both beams collinearly  before the setup and introducing them into the spectrometer. Finally, we note that with additional holes drilled into selected optics,  a pump beam to photoexcite the sample for pump-probe measurements could be introduced to the setup.

\begin{acknowledgments}
The authors thank Edward Kaczanowicz for assistance with custom designed polarizer mounts for the stepper motor stage. DHT acknowledges Temple University start-up funds.
\end{acknowledgments}

\bibliography{SHGFD}

\end{document}